\documentclass[manuscript, screen, acm]{acmart}
\AtBeginDocument{%
  \providecommand\BibTeX{{%
    \normalfont B\kern-0.5em{\scshape i\kern-0.25em b}\kern-0.8em\TeX}}}

\setcopyright{acmcopyright}
\copyrightyear{2023}
\acmYear{2023}
\acmDOI{}
\acmConference[GenAICHI 2023]{Generative AI and HCI at CHI 2023}{April 28, 2023}{Hamburg, Germany}

\begin{document}

\title{Exploring outlooks towards generative AI-based assistive technologies for people with Autism}

\author{Deepak Giri}
\email{dgiri@iu.edu}
\affiliation{%
  \institution{Indiana University-Purdue University Indianapolis}
  \country{United States}
}

\author{Erin Brady}
\email{brady@iupui.edu}
\affiliation{%
  \institution{Indiana University-Purdue University Indianapolis}
  \country{United States}
}

\renewcommand{\shortauthors}{Giri and Brady}

\begin{teaserfigure}
  \includegraphics[width=\textwidth]{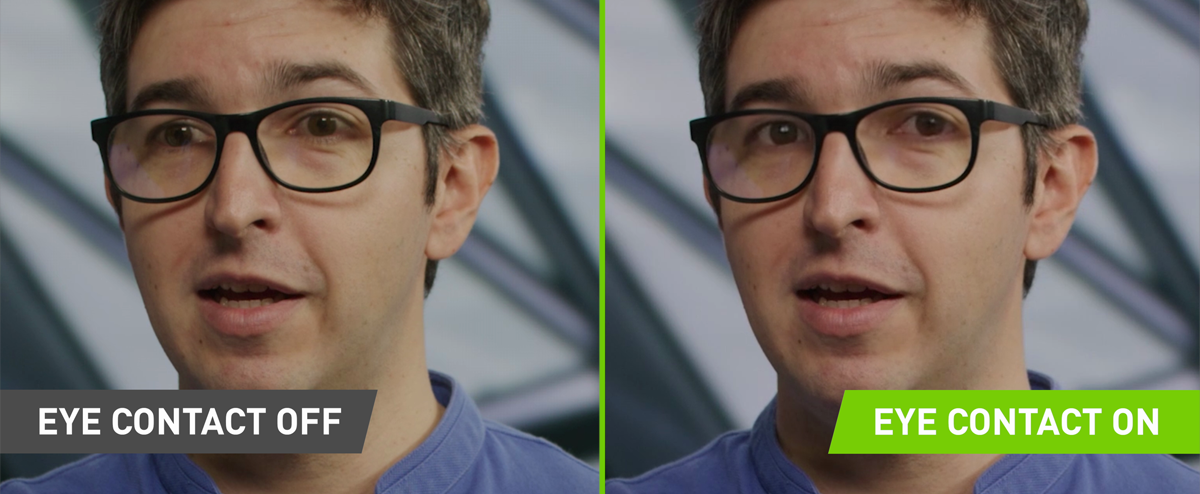}
  \caption{NVIDIA Broadcast 1.4 Eye Contact functionality, 2023.}
  \Description{NVIDIA Broadcast 1.4 Eye Contact functionality OFF and ON}
  \label{fig:teaser}
\end{teaserfigure}

\begin{abstract}
  The last few years have significantly increased global interest in generative artificial intelligence. \emph{Deepfakes}, which are synthetically created videos, emerged as an application of generative artificial intelligence. Fake news and pornographic content have been the two most prevalent negative use cases of deepfakes in the digital ecosystem. Deepfakes have some advantageous applications that experts in the subject have thought of in the areas of filmmaking, teaching, etc. Research on the potential of deepfakes among people with disabilities is, however, scarce or nonexistent. This workshop paper explores the potential of deepfakes as an assistive technology. We examined Reddit conversations regarding Nvdia’s new videoconferencing feature which allows participants to maintain eye contact during online meetings. Through manual web scraping and qualitative coding, we found 162 relevant comments discussing the relevance and appropriateness of the technology for people with Autism. The themes identified from the qualitative codes indicate a number of concerns for technology among the autistic community. We suggest that developing generative AI-based assistive solutions will have ramifications for human-computer interaction (HCI), and present open questions that should be investigated further in this space.
\end{abstract}



\maketitle

\section{Introduction}
Deepfakes are hyper-realistic content (commonly videos) created by generative artificial intelligence \cite{westerlund_emergence_2019}. Traditionally, people have been manipulating videos through editing tools and these kinds of videos are called cheapfakes \cite{shahid_it_2022}. Though cheapfakes are easier to detect by anyone, deepfakes on the other hand are so novel in nature that it is harder to identify even with digital forensic techniques. People have come a long way in creating deep fakes since the introduction of Generative Adversarial Networks \cite{goodfellow_generative_2014}, as they now only need smaller training datasets to create convincing content. Although deepfake detection will not be covered in this work, researchers predict that it will eventually call for more robust and dynamic algorithms \cite{passos_review_2022,harwell_top_2019}.

Deepfakes are notoriously used to generate pornographic content and it makes literally anyone a potential target \cite{noauthor_fake-porn_nodate}. Deepfakes are not limited to producing personal harm, they can destabilize governments, economies and specific target groups \cite{vaccari_deepfakes_2020, westerlund_emergence_2019,noauthor_deepfake_2019}. While deepfake's bad applications are readily apparent, experts in the field and researchers are also seeing certain advantages of the technique. Deepfake has been successfully used by filmmakers to transform older actors into their youthful selves \cite{noauthor_how_nodate}. Artificially created characters can be used in Deepfake as a teaching tool to actively engage students in the classroom \cite{kosmyna_thinking_2020,noauthor_male_nodate}. Moreover, deepfakes have been suggested as potential digital assistive technologies for individuals who have lost their vocal or facial expression due to an injury or illness \cite{pataranutaporn_ai-generated_2021}. Research on human perception \cite{tahir_seeing_2021,wohler_towards_2021}, attitudes \cite{shahid_it_2022}, and dialogues \cite{gamage_are_2022} about deepfake videos have previously been conducted. Studies investigating the perceptions and attitudes of people with disabilities toward prospective deepfake-based assistive technologies do not exist, though.

One specific area where a new deepfake technology has been proposed as beneficial for disabled users is in making eye contact over video calls. Making direct eye contact with someone is a social behaviour which has emotional and communicative meaning \cite{senju_eye_2009}. Eye contact is recognized as an important social signal, indicating that the speaker intends to communicate with someone, and can lead to emotional responses which may influence how the speaker’s message is interpreted \cite{senju_eye_2009}. However, people with autism have difficulty maintaining eye contact during conversations \cite{senju2009atypical}, to the point that “abnormalities in eye contact” are explicitly included as part of the DSM-5 diagnostic criteria for autism. For people with autism, prolonged eye contact feels invasive, and can elicit negative emotional and physiological responses or trigger sensory overload \cite{trevisan_how_2017}.

The introduction of the eye contact functionality by the semiconductor behemoth Nvidia in its conferencing software, broadcast v1.4, has sparked some discussion within the autistic community. This technology artificially modifies live video content to make it appear as if the person in the video frame is making direct eye contact with the viewer, even if in reality their eyes are focused elsewhere. While the company does not frame their eye contact technology as solely an assistive technology, they highlight this as a valuable application of the feature, mentioned in Nvidia Technical Blog (2023):  \emph{"Maintaining eye contact is also occasionally problematic for many physiological reasons. A lot of kids and adults find it difficult to establish and maintain eye contact.”} \cite{noauthor_improve_2023}

Prior research within the HCI community has sought to improve eye contact behaviours specifically among people with autism, through interventions such as wearable glasses which track gaze and give feedback when the wearer is not maintaining eye contact \cite{kline_superpower_2019}, or eye tracking games which incentivize maintaining eye contact \cite{ng_recommending_2018}. Other HCI work has critiqued these types of technologies for attempting to make autistic people conform to neurotypical conversational norms \cite{williams_perseverations_2020} and identified ways in which autistic people may try to “mask” their autism and appear neurotypical through behaviours like faking eye contact on a video call \cite{zolyomi_managing_2019}. These masking behaviours require significant cognitive effort for autistic people, and autistic people appreciated that making direct eye contact is not a conversational norm on video conferencing tools \cite{zolyomi_managing_2019}. It is unclear from this whether people with autism would find artificially generated “eye contact” beneficial, neutral, or harmful.

In this study, we examined Reddit comments on posts related to Nvidia's technology to gain early insights into people’s reactions to generative AI-based assistive technologies. The eye contact function from Nvidia was released in early January 2023, and since then 162 relevant comments have been gathered. We conducted qualitative coding to uncover the ideas behind the conversations and used these themes to identify open questions in the space of generative AI-based assistive technologies.

\section{Data and findings}
We manually identified 9 Reddit discussion threads where users were discussing Nvidia’s eye contact functionality in relation to Autism through a web search for relevant keywords (e.g., “Nvidia autism site: reddit.com“). 5 of these discussions were in the /r/autism subreddit, and the others were from /r/AutisticAdults, /r/autismmemes, /r/nextfuckinglevel, and /r/nvidia.  We collected all comments from the posts in autism-specific subreddits. For posts in other subreddits, we manually screened the comments and only collected the ones which explicitly discussed autism or neurodiversity. From these posts, we collected 162 comments. 

The lead researcher did an initial open coding pass on the comments. The researchers then met to discuss the codes and expand the codeset further.  Data were then organized into high-level themes, described in detail below. Additionally, we recorded whether the commenters self-disclosed in their comments that: (a) they themselves had autism (20 commenters); (b) they did not have autism, but had close family members or friends with autism (4); or (c) they did not have autism (22).  The other 116 comments did not include any disclosure of disability status.

\subsection{Sentiment towards prospective generative AI assistive technologies}

A majority of comments (N=88) expressed negative responses towards Nvidia's eye contact feature. A large portion of Redditors angrily criticized the feature, describing it as “nightmare fuel” and indicated that they would choose not to use it at all. Many people criticized it as a "masking strategy" that wouldn't improve their real-world circumstances, or questioned the value of eye contact for communication at all. Comments described concerns both about using the feature to alter their own videos, and about having to be on video calls with others using the feature.

Despite these concerns, a considerable number of comments (N=46) were in support of the functionality. Autistic individuals described how the feature could be utilized to help them livestream effectively, confidently communicate with neurotypical individuals, and succeed in jobs which are communication-intensive like sales. Some users had even higher expectations for this technology, wondering if it could eventually be used for in-person encounters (e.g., “\emph{Do they gave special glasses that made that in real life? I'm autistic}”). Others sought some kind of control over the functionality. Neurotypical people made 14 of the 46 comments in favour of the feature because they thought it would make people on the spectrum more at ease during online encounters. In the comments about the feature's usefulness, 3 out of 46 Redditors mentioned relatives who had autism.

Among the 46 comments where commenters self-disclosed whether they were neurotypical or autistic, there was much more disparity in the opinions of the commenters who were autistic (10 positive, 8 negative, 2 neutral/no opinion) than those of commenters who identified as neurotypical (14 positive, 4 negative, 4 neutral/no opinion).

\subsection{Video conferencing behaviour and setup}

The comments revealed the video-conferencing behaviour of people with autism. People stated that turning off their cameras helps them with their social anxiety. Some commenters admitted their tendency of looking at themselves in the video conferencing software and anticipated that the functionality would increase this behaviour. The foundational idea of the eye contact functionality was criticized in one of the comments "\emph{most individuals don't look at their camera since they're too busy staring at their screen.}" There have been questions about whether this feature will be useful to everyone depending on their workstation configuration. An individual quoted “\emph{The device camera isn't in the same location as people's eyes, which results in nobody actually looking like they're looking at other people's eyes.}” Another comment mentioned that the eye contact feature would make things worse when people have a multiple monitor setup. The educator mentioned that eye movement helps their students know when they are referencing the slides in the second monitor. A game streamer questioned the usefulness of this feature in his line of work, as it would make things strange for him because it would appear to his audience as though he isn't paying attention to the game at all. 

\subsection{Relevance of eye contact}

During our analysis, we found out that a significant number of comments (N=50) expressed eye contact to be irrelevant in social interactions. As evidenced by an individual who claims to feel less apprehensive when interpreting facial expressions and tone of voice as opposed to direct eye contact. Some comments harshly criticized the need for eye contact and had a strong stance that generative AI should not normalize it. One comment quoted “\emph{I don’t want to be looking into the camera and I don’t want software doing it for me either. It would feel like I’m being forced to make eye contact against my will even if I’m not physically doing so.}” Another set of comments questioned the societal aspirations of neurotypical individuals and felt the eye contact functionality was a personal attack on their social interaction preferences. Many comments advocated for a need of spreading awareness in society in the sense that it is completely normal to not make eye contact. One comment elaborated “\emph{Basically, making other people less uncomfortable is the wrong message. Better is to normalize the idea that not everyone is “normal” and that things like eye contact shouldn’t be the social expectation or demand. That’s my opinion anyway. Other people have differing opinions, but I really don’t think conformity should be the goal.}” While most comments were direct disapproval of eye contact in social interaction, some comments looked at it through the perspective of culture and neurotypical individuals. One comment stated that the relevance of eye contact would have stemmed from the societal belief that it is a sign of attentive listening and sincerity when speaking. While another comment mentioned the relevance of eye contact in certain cultures and is not necessary to be mimicked everywhere. 

\subsection{Interactions are creepy}
Commenters (N=35) felt the eye contact functionality to be creepy and disturbing. One comment quoted “\emph{This would only make me look insane and murderous.}” Others also pointed out that the eye contact function was strange because social interactions don't typically entail long stretches of eye contact; instead, it appears more like gazing. Those with autism noted that it was not only ineffective but would also increase people's lack of confidence in one another, which would have long-term effects on society. Many believed that the gaze intensity was one of the things that made eye contact unsettling. One commenter brought up the possibility of using technology in scripted media. Some also expressed privacy concerns since they thought using the software would make them feel watched over.

\section{Discussions}
This preliminary investigation qualitatively examined public sentiment towards generative AI-based assistive technologies for people with autism. Our findings suggest that there are mixed opinions towards generative AI-based assistive technology among both autistic and neurotypical individuals. Although we observed a negative inclination towards the technology, there were also significant comments regarding its perceived usefulness, especially from neurotypical commenters. These findings give us an initial look at public sentiment towards the use of generative AI for assistive purposes, but need to be validated through interviews and contextual inquiry in future work. Below, we present some open questions raised by our analysis, which we hope to discuss with the other workshop attendees and continue to explore in future work.

\textbf{Usability of Generative AI-based Technologies:} We came across various comments questioning the usability of generative AI-based assistive technology. The comments described how the technology would not embed well with their workstation setup, nature of work and behaviour. This calls for human factors research to better understand how autistic individuals would like to interact with generative AI-based assistive technologies and how their environment would support the same.

\textbf{Challenging Neurotypical Communication Norms:} Many commenters pushed back against the ableist assumptions embedded into this style of generative AI systems, questioning the necessity of making eye contact during communication. Generative AI systems are trained on large datasets, but these datasets often represent the experiences of people without disabilities.

\textbf{Investigating Uncanniness:} The unsettling and unnatural nature of this potential assistive technology based on generative AI was one of the key themes that came out of our work. The comments exemplified the technology's position in the uncanny valley, which is characterized by a nonlinear relationship between affinity and perceived human likeness \cite{mori_uncanny_2012}. Researchers have been attempting to comprehend this phenomenon with AI \cite{avdeeff_artificial_2019, weisman_face_2021}, and it gives us a direction to comprehend it specifically for assistive technologies based on generative AI to produce reliable interventions. Additionally, prior research in the accessibility domain indicates that people may react more positively to new technologies if they know they are being used for assistive purposes \cite{profita2016effect}; it would be useful to explore whether neurotypical people have more positive reactions to generative AI-based assistive technologies if they know why they are being used.

\bibliographystyle{acm}
\bibliography{Exported-Items}

\appendix
\end{document}